\documentclass[12pt]{article}  

\usepackage{amsmath,amsfonts,amssymb}
\usepackage{graphicx}
\usepackage{float}
\usepackage[colorlinks=true, allcolors=blue]{hyperref}
\usepackage{makecell}

\providecommand{\keywords}[1]
{
  \small	
  \textbf{\textit{Keywords---}} #1
}

\title{HARMONI at ELT: tolerance analysis and expected as-build imaging performance of the infrared spectrograph}
\author{Eduard Muslimov$^{a,b}$, Edgar Castillo-Domínguez$^{a}$, James Kariuki$^{a}$,\\
Jorge Chao-Ortiz$^{c}$, Matthias Tecza$^{a}$, Elliot Meyer$^{a}$,\\
Zeynep Ozer$^{a}$, Fraser Clarke$^{a}$, Niranjan Thatte$^{a}$\\
on behalf of the HARMONI consortium
\\
\small a--Department of Physics, University of Oxford, \\
\small Keble Rd, OX14 3RH Oxford, UK\\
\small b -- Aix Marseille Univ., CNRS, CNES, LAM, Marseille\\
\small c--Indra Sistemas, S.A., Madrid, \\
\small Avenida de Bruselas N 35, 28108 Alcobendas, Spain\\
}

\begin{document} 
\maketitle{}

\begin{abstract}
HARMONI is the first light visible and near-IR integral field spectrograph for the ELT. It covers a large spectral range from 470nm to 2450nm with resolving powers from 3300 to 18000 and spatial sampling from 60mas to 4mas. It can operate in two Adaptive Optics modes - SCAO (including a High Contrast capability) and LTAO - or with NOAO. The project is preparing for Final Design Reviews.
The integral field spectrograph is a key sub-system of HARMONI instrument, which forms the 2D spectral image and projects it onto the scientific detector. It has 40 operational modes with different platescales and gratings covering the band of 811-2450 nm with three resolution grades. In each of this configurations the as-built spectrograph wavefront error is strictly limited.
We perform the sensitivity analysis for measurable and unknown errors and build the errors budget on this basis. Then we correct the values for the actual technological limits and perform a three-stage Monte-Carlo analysis combined with simulation of a few specific effect as the holographic grating wavefront error. Eventually, we show that it is possible to reach the target image quality in terms of the wavefront error and spectral resolution for the entire sub-system with practically feasible tolerances on  design parameters.
\end{abstract}


\section{INTRODUCTION}
\label{sec:intro}  
HARMONI is a near infrared and optical integral field spectrograph (IFS), it will be installed on one of the Nasmyth ports of the Extremely Large Telescope (ELT) on top of Cerro Armazones. The IFS is the primary HARMONI system and is responsible for the extraction of the spectra from the science beam delivered to the IFS focal plane. In turn, the IFS is split into a few sub-systems. 

The IFS will slice a single contiguous field measuring 206 x 152 spaxels (“SPAtial piXELS”) in size to form a slit, which in turn will be dispersed across the slit width to form spectral images at four different scales. The four spaxel scales are $60 \times 30$, $20 \times 20$, $10 \times 10$ and $4 \times 4 mas$  with nominal fields of view on the sky of $9.12 \times 6.42", 4.28 \times 3.04", 2.14 \times 1.52"$ and $0.85 \times 0.61"$ with increasing spatial resolution respectively. The spectral resolving power settings of $R=3700, 7900$, and $19000$  will be provided in all spatial scales, covering the wavelength range  from $0.811$ to $2.469 \mu m$ \cite{Thatte22}.

The last of the optical sub-systems is marked as the IFS Spectrograph subsystem (ISP) and it performs such key  functions as dispersing the light and creating the spectral image on the science detector. The spectrograph design is driven by the volume restrictions, throughput requirements and the cryogenic operation conditions. The current baseline optical design of this sub-system is shown in Fig.~\ref{fig:layount} below. Each of the 4 ISP's installed in HARMONI will consist of:
\begin{itemize}
    \item the three-mirror anastigmate (TMA) type collimator with focal length of $1546.4 mm$ and maximum $f/\#=11.71$, covering the wide linear field of $-270.8.. +270.8 mm$ in the spatial direction,
    \item fold mechanism mirrors (FMM) used to change the angles of incidence for different resolution settings, 
    \item 10 interchangeable transmission volume-phase holographic (VPH) gratings organized in 3 families - low, medium and high resolution (LR, MR and HR, respectively) and mounted on a wheel mechanism (infrared gratings mechanism - IGM),
    \item Infrared camera module (ICAM) with  focal length of $349.6 mm$ and maximum $f/\#=2.67$, working in the entire band from $0.811$ to $2.45 \mu m$.
\end{itemize} 

The visible channel shown here and having its' own folding mirror, grating and camera lens will be installed in 2 out of 4 ISP's and its' design is developed in parallel. Note also that the science detector with its' electronics and mechanisms represents a separate sub-system in the instrument structure.

   \begin{figure} [ht]
   \begin{center}
   \begin{tabular}{c} 
   \includegraphics[width=0.85\textwidth]{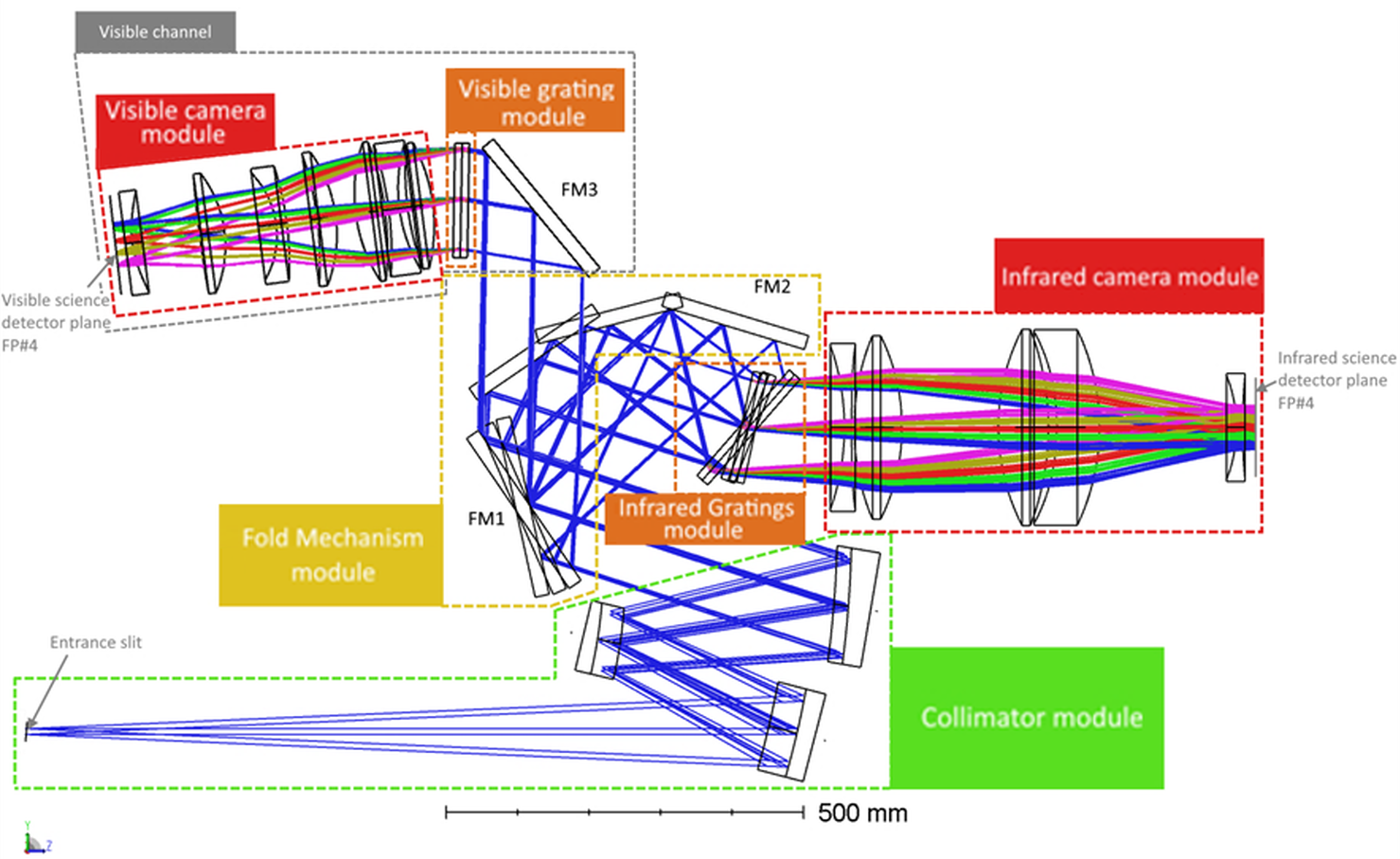}
   \end{tabular}
   \end{center}
   \caption[lay] 
   { \label{fig:layount} 
General view of the integral field spectrograph optical system design.}
   \end{figure} 

\begin{figure} [ht]
   \begin{center}
   \begin{tabular}{c} 
   \includegraphics[width=0.95\textwidth]{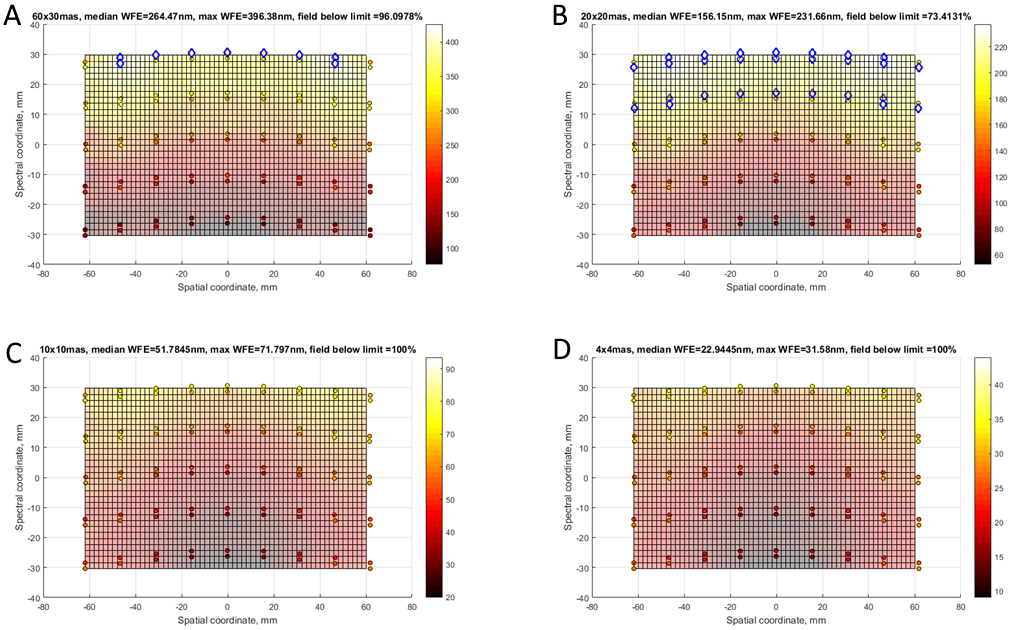}
   \end{tabular}
   \end{center}
   \caption[nominal] 
   { \label{fig:nominal} 
Wavefront error maps for the worst-case configuration in the nominal ISP sub-system design: A -- 30x60 mas, B -- 20x20 mas, C -- 10x10 mas, D -- 4x4 mas.}
   \end{figure}

There is a number of challenges associated with the ISP design, such as operating in vacuum under $T=130K$ and providing high performance over the wide spectral range and large spatial field with each combination of grating and spaxel scale settings. In addition to this, there is a task of ISP integration to the instrument, which leads to using a combination of different metrics to define the  required performance and creating of a complicated break-down for each of them. In particular, the image quality is estimated through each module by the wavefront error (WFE) computed as RMS over the pupil and then defined for the $95\%$ of best spaxels in the image. In parallel, the quality is controlled via the full width at half maximum (FWHM) of the slitlet monochromatic image and, further, directly via the actual spectral resolving power, which incorporates the magnification and linear dispersion. Fig.~\ref{fig:nominal} exemplifies the WFE maps computed for the nominal baseline design with MR2 grating. This is the most risky configuration - see the datapoints marked with blue diamonds, which approach the WFE limit.  

Each of this metrics depends on a large number of design parameters as radii, thicknesses and material properties etc.  In combination with the tight performance requirements this leads to high sensitivities to the manufacturing and alignment errors. So, for the HARMONI sub-systems the standard stage of sensitivities analysis and tolerances definition becomes less conventional as it requires  some scrupulous analysis with custom tools and may challenge the current technological level in a few ways.

In the present paper we discuss the tolerancing process for ISP-NIR sub-system, explain our approaches and demonstrate the analysis results. In general, we first consider calculation of individual sensitivities and definition of corresponding tolerances with a simple root square sum rule. This is referred to as a direct problem. Then, after correcting the computed tolerances according to the actual technological capabilities, we solve the inverse task, i.e. define the expected performance of as-built sub-system with all the errors. These errors, in turn, fall into two categories: the first represents simple geometrical deviations and position misalignments or surface shape errors, which could be relatively easily modelled with ray-tracing in a standard Monte-Carlo analysis loop; the second contains more complex effects related tot the materials properties, diffractive optics etc, which have to be modelled separately. Then the effects are combined to derive the final conclusion on the expected performance. The rest of the paper is organized according to this general analysis sequence.

\section{Direct problem - tolerances definition}
\label{sec:direct}  

The tolerance analysis is performed at the module level, using the wavefront error, pointing and magnification requirements breakdown. For every module, sensitivity tables are generated by introducing individual small deviation and calculating teh numerical derivatives under different alignment scenarios. Using these sensitivities, the tolerances on individual parameters are generated with a root square sum (RSS) rule. 
For the collimator module this analysis includes:
\begin{itemize}
    \item Linear displacement of each mirror with respect to the mechanical pivot point $ \Delta x, \Delta y, \Delta z$,
    \item Rotation of each mirror with respect to the mechanical pivot point $\alpha_x, \alpha_y, \alpha_z$. As the mirrors are off-axis aspheress and they are aligned with respect to the offset pivot different from the surface vector, the design is not invariant with respect to the $\alpha_z$ clocking angle,
    \item Surface shape error in low spatial frequency $\Delta z_{Zern}$ domain represented by Zernike polynomials, modes 5-21. The Zernike modes are added with centring to the aperture centre, not the vertex, therefore they are generated one-by-one,
    \item Defocus, generated and introduced as Z4 Zernike mode,
    \item Mid-spatial frequency surface errors introduced separately for the X and Y directions at a few probing periods, which is expectable for a highly rectangular aperture. Each error pattern represents a simple sine wave. Some details on their modelling are discussed below in Sec.~\ref{sec:inverse}. 
\end{itemize}
	
All of the calculations are performed for 12 field positions (6 in \textit{X} direction, 2 in \textit{Y} direction), which cover the full field uniformly. The design is fully achromatic, so in the sake of simplicity all of the computations are made for $\lambda=1000 nm$. The main criterion is the maximum RMS WFE over the field, as it’s defined before. 
The analysis is performed in 2 cycles. First, the sensitivities are computed without any compensators. The second cycle uses compensators. The full set of compensators include $\Delta x, \Delta y, \Delta z$ linear movements at \textit{M2} and \textit{M3} mirrors, while \textit{M1} is used as the reference. We presume that the coordinates for alignment will be measured by a coordinate-measuring machine (CMM). 
 
All of the parameters are classified into manufacturing, alignment and measurement errors. The manufacturing parameters relate to the surface shape error and it is supposed that they are known from the optical elements testing to a high precision, so compensation can be used for these errors. As the testing data contain all of the errors, all of them will be substituted to the model to compute the compensators simultaneously (e.g. as a surface or phase error computed from the interferogram) regardless of the expected compensation efficiency. The residual errors after the alignment and measurements (for the compensators) are unknown, so the for them the uncompensated scenario applies.
The calculated sensitivities for every parameter P  represent a numerical derivative, computed as
\begin{equation}
\label{eq:deriv}
\frac{\partial WFE_{max}}{\partial P} \approx\frac{max(WFE_{max}(+\Delta P),WFE_{max}(-\Delta P))}{\Delta P} ,
\end{equation}

where $\Delta P$ is a probing perturbation of the parameter, which roughly corresponds to the middle of the expected error margin and the $ WFE_{max}$ is the corresponding degradation of the max WFE across the field. So, the sensitivities always presume the direction of steepest criterion change and ignore the dependence non-linearity.
For the defocus component the following formula is used to convert it to the local radius of curvature
\begin{equation}
\label{eq:defoc}
\frac{\partial WFE_{max}}{\partial R} =\frac{\partial WFE_{max}}{\partial Z_4} 4\sqrt{3} Z^2_4 .
\end{equation}

The low-frequency surface shape errors are recomputed to the RMS error via root square sum

\begin{equation}
\label{eq:zern}
\frac{\partial WFE_{max}}{\partial RMS} =\frac{ \sqrt {WFE^2_{max,k}}}{\sqrt{\Sigma Z^2_k}}.
\end{equation}

For estimation of the reachable tolerances on individual parameters we used the vlaues known from such publications as.
The full set of results is not shown here, but the key figures found there are as follows:
\begin{itemize}
    \item The overall maximum WFE based on root square sum rule is \textit{85.60, 50.90, 46.58} and  \textit{39.55 nm} for the 3\textit{0x60, 20x20, 10x10} and \textit{4x4 mas} scales, respectively if the high  precision level is applied to each of the errors;
    \item The corresponding contribution of alignment errors are \textit{26.84, 11.38, 3.59} and \textit{1.50 nm};
    \item All of the required tolerances are above the known technological limits.
\end{itemize}

The analysis of sensitivities for the ICAM is very similar. The target image quality can be reached only with compensators. For some parameters, the actual values including the errors can be measured and reported by the manufacturers. Then these values can be substituted into the optical system model. After optimization it will be possible to implement the found best axial lens position by grinding the shims connecting each of the lens mounts to the common housing and introduce the required axial shift, tip and tilt angles of the detector (“D+L” mode). Other group of errors cannot be measured exactly and will be compensated in an assembled camera by linear and angular movement of the detector (“D” mode).  As in the previous case, the parameters errors are classified into manufacturing errors, alignment errors and measurement errors, depending on the stage of the production process, on which they occur.

In comparison with the collimator, the ICAM design demonstrates a higher sensitivity:
\begin{itemize}
    \item The expected WFE found with RSS rule are \textit{358.9, 200.8, 65.4} and	\textit{29.2 nm} for the four spaxels scales, respectively;
    \item The required tolerances on the most of surface and element tilts are higher than the ones corresponding to the typical precision level.
\end{itemize}

For the gratings module and the fold mirrors the tolerances are driven by the budget breakdown and the technologically feasible limits declared by the suppliers, so the tolerances are applied directly and not considered here. 
For the multi-element modules, i.e., the collimator and camera, the tolerances are derived from the required WFE using the sensitivities computed above. The procedure is similar for both of the modules.
First, the allowable WFE degradation is defined. For the collimator we presume that the overall error $WFE_{coll, tot}$ represents a linear sum of the as-designed (nominal) one $WFE_{coll,nom}$ and the WFE introduced by all of the errors. The analysis is performed for the \textit{60x30 mas} platescale as it has the highest aberrations.
 For the camera, the as-designed WFE is higher than the allocated budget for some configurations. This means that the initially allocated WFE budget can’t be applied as a criterion. Instead, we use a simplistic criterion, allowing $10\%$ degradation of the RMS WFE median across the 2D field (spectral and spatial). As the camera works in multiple configurations, we carry out the calculation for the one, which has the worst as-designed performance, namely \textit{60x30 mas} platescale with \textit{MR2} grating.  This allows to avoid simulating the global optimum search each time and provides a conservative estimation.    
The resultant allowable WFE degradation is then corrected for the measurement errors $WFE_{coll,meas}$. It represents an unknown random value, so the RSS rule is used. The remaining WFE is uniformly distributed across the \textit{N} manufacturing and alignment errors using the RSS rule again.
So, the WFE contribution per parameter \textit{P} is calculated as

\begin{equation}
\label{eq:wfe_collP}
\Delta WFE_{coll,P}=\sqrt{\frac{(WFE_{coll,tot}-WFE_{coll,nom})^2-WFE^2_{coll,meas}}{N_{coll}}},
\end{equation}

\begin{equation}
\label{eq:wfe_camP}
\Delta WFE_{cam,P}=\sqrt{\frac{0.1WFE^2_{cam,nom}-WFE^2_{cam,meas}}{N_{cam}}} .
\end{equation}

As the allowable WFE degradation per parameter is defined, the corresponding parameter increment is found via the numerical derivatives 
\begin{equation}
\label{eq:dp}
\Delta P =\frac{\Delta P}{\partial WFE_P/ \partial P}.
\end{equation}

The values are compared against the standard level of precision \cite{Lane2012}. If the increment larger than the standard tolerance, it is truncated down to the standard value, the WFE contribution is updated as

\begin{equation}
\label{eq:trunc}
\Delta WFE_{loose,P} =\Delta P_{stand} \frac{\partial WFE_P}{\partial P},
\end{equation}

 and the released WFE budget is re-distributed among the rest \textit{K} parameters

 \begin{equation}
\label{eq:redist1}
\Delta WFE'_{loose,P} =\sqrt{\frac{(WFE_{coll,tot}-WFE_{coll,nom})^2-WFE^2_{coll,meas}-\Sigma WFE^2_{coll,loose}}{K_{coll}}},
\end{equation}

\begin{equation}
\label{eq:redist2}
\Delta WFE'_{cam,P}=\sqrt{\frac{0.1WFE^2_{cam,nom}-WFE^2_{cam,meas}-\Sigma WFE^2_{cam,loose}}{N_{cam}}} .
\end{equation}

Then the individual increments for the \textit{K} remaining parameters are computed again with (\ref{eq:dp}) and accepted as the tolerances.
For some specific types of tolerances, the entire analysis should be repeated for all the plate scales. This applies to the surface shape irregularities at low spatial frequency to generate the surface root mean square (RMS) and peak-to-valley (PtV) errors tables, and at mid frequencies to generate the power spectrum distribution (PSD).
These tolerances are used to issue the drawings, which are sent to potential suppliers for the evaluation. As the suppliers’ responses are summarized, the expected errors margins are corrected by setting technologically feasible bottom limits. The errors margins obtained this way are used for Monte-Carlo simulations of the as-built expected performance, which are presented in the section below.

\section{Inverse problem - expected performance}
\label{sec:inverse}  

\subsection{Conventional optics performance}
\label{sec:conv}

The core simulation used to predict the as-built performance is Monte-Carlo analysis in a raytracing model. It incorporates all the known tolerances defined by manufacturing, alignment and measurements for the lenses, mirrors and plates. Since the nominal performance leaves little margin in comparison with the requirements and the tolerances above estimated with the RSS method  are tight, we use a bottom-up approach for further analysis. This implies that the required tolerances were examined by the responsible ones and were either approved, or replaced by the closest technologically reachable value. 
With respect to the applicable compensators, the alignment and manufacturing errors can be classified into 3 groups. Therefore, we perform the Monte-Carlo simulations in three stages to reflect the integration process:
\begin{enumerate}
    \item We introduce the manufacturing errors, which may be measured with a high precision prior to the beginning of modules and sub-system integration. This group includes radii of curvature and thicknesses of individual optical components, low-frequency surface shape errors and the wedgeness on spherical surfaces. In this case all of the available compensators can be computed in advance and introduced into the model. The first stage includes a few loops:
    \begin{itemize}
        \item Random deviations for the collimator mirrors radii and low frequency surface shape errors (Zernike modes $Z_{5-21}$) are introduced. The distance from the entrance slit to \textit{M1} is adjusted to restore focusing for the median wavelength. Then the entire collimator is optimized for the overall WFE at the largest platescale, using the \textit{X,Y} and \textit{Z} compensators at \textit{M2} and \textit{M3} as defined before. The WFE is defined as RMS over the pupil and median over the spectral and spatial positions and configurations.
        \item Similarly, deviations in radii and surface shape ($Z_{4-37}$) of \textit{L2}, \textit{L3} and \textit{L5} lenses are introduced into the ICAM module. The focus at median wavelength is restored and then the lens airgaps and the detector axial position, tip and tilt are optimized for the median RMS WFE. 
        \item These simulations are run in a loop to generate 100 Monte-Carlo instances with random deviations of the specified parameters, acting simultaneously. Each instance is saved as a separate \textit{.ZMX} model file.
    \end{itemize}
    \item We open each of the files generated at stage 1 and introduce the manufacturing and alignment errors, which can’t be measured directly in advance, but their effect can be observed at different alignment stages and compensated accordingly. This includes finite alignment precision of the optical elements alignment in all 6 degrees of freedom (where applicable), alignment of modules with respect to each other, static and repeatable components of the positioning errors of the FMM and IGM mechanisms, gratings line densities and wedgeness, optical materials properties (refraction index, Abbe number, CTE), tilts and decenters of the aspheric surfaces in lenses and all kinds of measurement errors. For each of the 100 instances 1 output ZMX file is generated, so the total number doesn’t change. For the second stage we use 3 loops with different merit functions:
    \begin{itemize}
        \item Restoring the focusing at \textit{MR2} central wavelength with the detector position,
        \item Restoring the spectral image centering by shifting the detector in \textit{X} and \textit{Y},
        \item Finding the overall optimum for RMS WFE with the tip, tilt and focus of the detector. 
    \end{itemize}
    \item We introduce the errors, which correspond to the moving parts of ISP and can’t be calibrated out with a static compensator. This includes the dynamic repeatability and stability errors of the FM1, FM2 and IGM. There are no optimization loops for these errors as there are no practical means to compensate them. Similarly, the Monte-Carlo instances are generated 1-to-1 for the stage 2 array.
\end{enumerate}
	
Another presumption taken in this analysis is a set of probability distributions used to generate the errors. By default we use Gaussian statistics with the error margins of $\pm 2 \sigma$ from the nominal. For some specific cases it was shown that the as-manufactured values follow different statistics and used these distributions: radii of curvature are distributed uniformly, lenses and plates thicknesses are distributed normally, but have a shifted median $\mu$, so the margins are  $+1.337 \sigma$ and $-2.673\sigma$ \cite{Kaufman14}.

 \begin{figure} [ht]
   \begin{center}
   \begin{tabular}{c} 
   \includegraphics[width=0.97\textwidth]{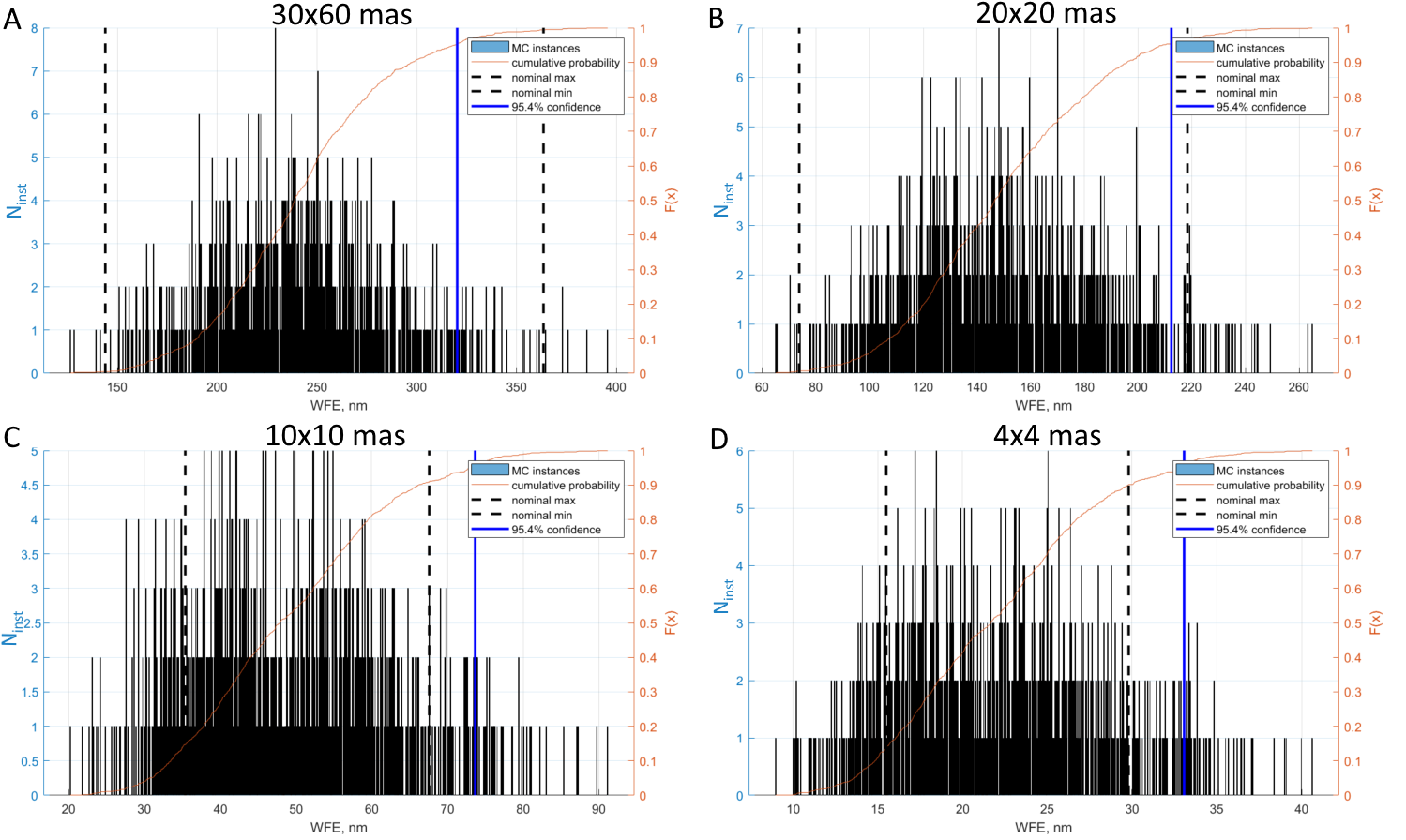}
   \end{tabular}
   \end{center}
   \caption[wfe] 
   { \label{fig:wfe} 
Monte-Carlo statistics for the wavefront error computed as a root-mean square over the pupil and 95-percentile over all the spaxels for 4 platescales: A -- 30x60 mas, B -- 20x20 mas, C -- 10x10 mas, D -- 4x4 mas.}
   \end{figure} 

The main outcome of this analysis is a statistical distribution of the ISP WFE defined via the 95-percentile metric. The distributions and cumulative probability plots for each of the 4 spaxel scales are shown in Fig.~\ref{fig:wfe}. In addition the WFE level corresponding to the target $95.4\%$ probability and the nominal values for the best and worst configurations are shown in the plots. One may note that the distributions are not perfectly symmetric having a widely dispersed part corresponding to the larger WFE. So, the well-known ratios of normal distribution can't be applied directly in this case.

In general, this analysis indicates that the additional WFE introduced by polishing and positioning of the mirrors and lenses is relatively small. The image quality at smaller platescales is expected to degrade more with respect to the as-designed performance. It should be also noted that as each grating change is considered as a separate instance, the overall WFE metric may be better than performance with a specific grating. This should be taken into account when interpreting this result and the subsequent conclusions.

\begin{figure} [ht]
   \begin{center}
   \begin{tabular}{c} 
   \includegraphics[width=0.65\textwidth]{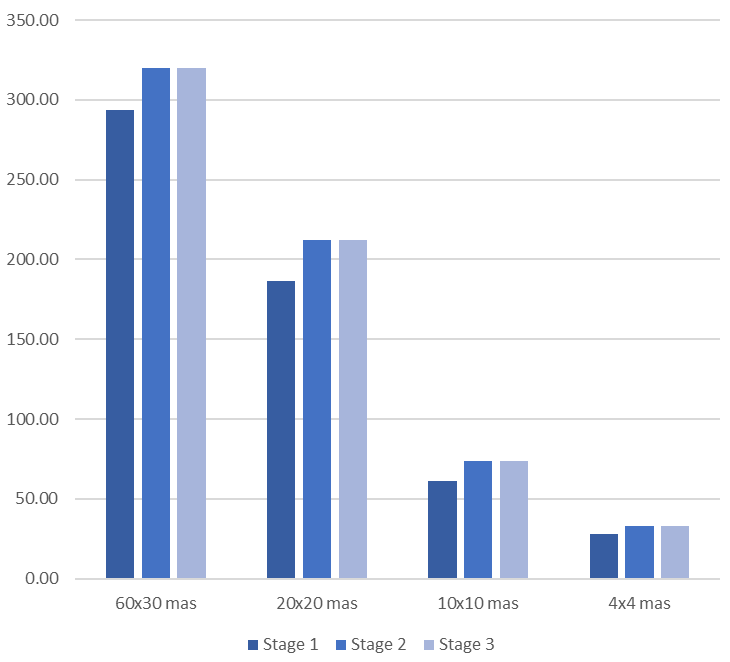}
   \end{tabular}
   \end{center}
   \caption[histo] 
   { \label{fig:histo} 
Change of the 95-percentile WFE per simulation stage.}
   \end{figure}

The influence of errors of each kind is illustrated in the diagram below (Fig.~\ref{fig:histo}). The first group of errors analyzed at stage 1 has the largest impact, while the residual dynamic errors at stage 3 have almost negligible influence and should be compared against the pointing error requirements rather than the WFE criterion. 

\subsection{Other factors and overall system performance}
\label{sec:overall}

There is a number of parameters, which would be difficult to calculate directly in complex simulations loops, so, they are taken into account elsewhere.

\begin{figure} [ht]
   \begin{center}
   \begin{tabular}{c} 
   \includegraphics[width=0.9\textwidth]{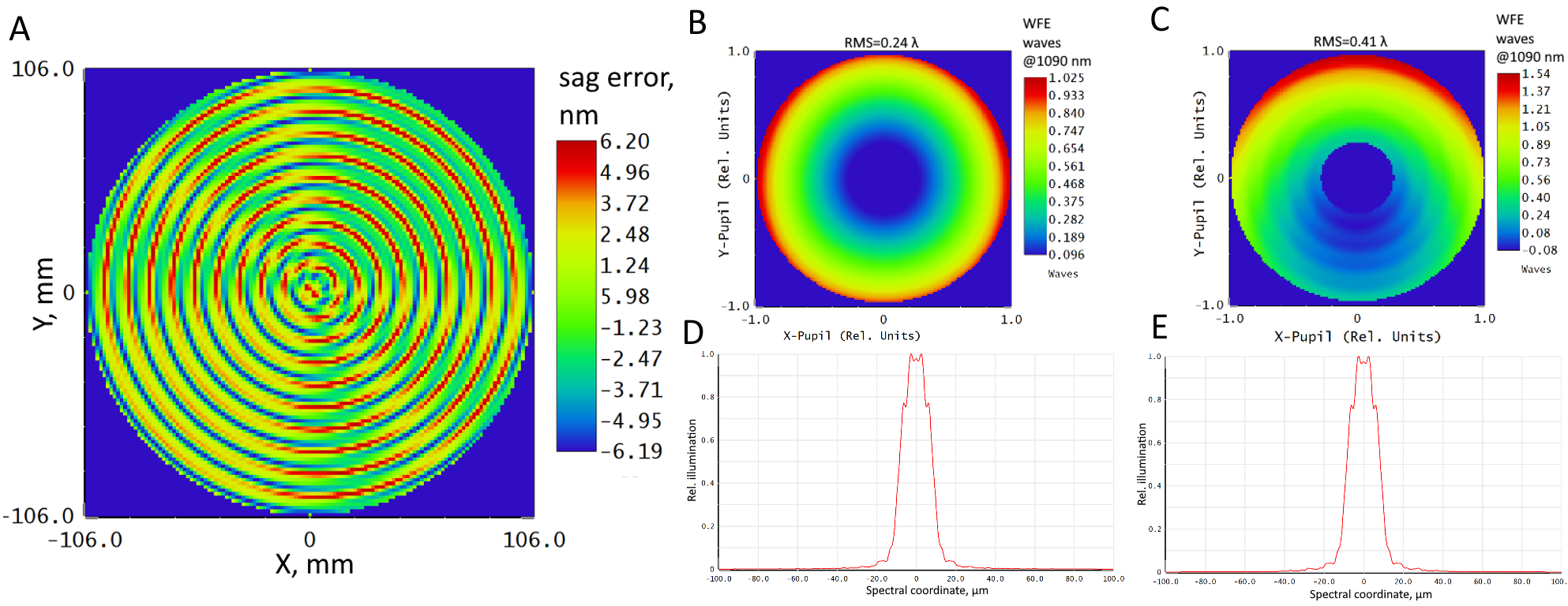}
   \end{tabular}
   \end{center}
   \caption[ripples] 
   { \label{fig:ripples} 
Mid spatial frequency toolmarks simulations: A -- surface error simulated with 2 carrier frequencies, B -- initial WFE, C -- WFE with the ripples at L11 surface (10 times exaggerated), D -- initial LSF, E -- LSF with the ripples at L11 surface (10 times exaggerated). }
   \end{figure}

Mid-frequency surface shape errors on lenses and mirrors are very likely to occur with diamond turning polishing and may even become the leading factor for the surface shape precision \cite{Aikens08}. They cannot be adequately represented by Zernike polynomials below the $231^{th}$ mode, which are used in $Zemax Optics Studio^{TM}$ by default. So they are  simulated using a dedicated script, introducing regular sinusoidal ripples with a few probing periods. This approach allows one to use explicitly the waviness definition from ISO10110-8 standard to specify the mid-frequency error at a few typical spatial periods, which may be more convenient than using a PSD, which is more suitable for picth polishing technology generating a  distributed spatial frequencies.  On another hand, it requires significant computational resources because of the increase surface sampling required. In Fig.~\ref{fig:ripples} the representation of such toolmarks and their effect is shown on the example of ICAM \textit{L1} left surface. The effect is barely visible, so the error is exaggerated from 6 to 60 nm PtV. It's useful to note that even a large error has virtually no effect on the LSF FWHM, but makes an impact on the WFE map \cite{Harvey21}. This means that the effect of the mid-frequency errors should be compared separately against the scattered light requirements later on. This simulation is difficult to run in  a loop, so further, we use RSS rule to add this effect to the WFE budget, but neglect it in the FWHM calculations.

\begin{figure} [ht]
   \begin{center}
   \begin{tabular}{c} 
   \includegraphics[width=0.9\textwidth]{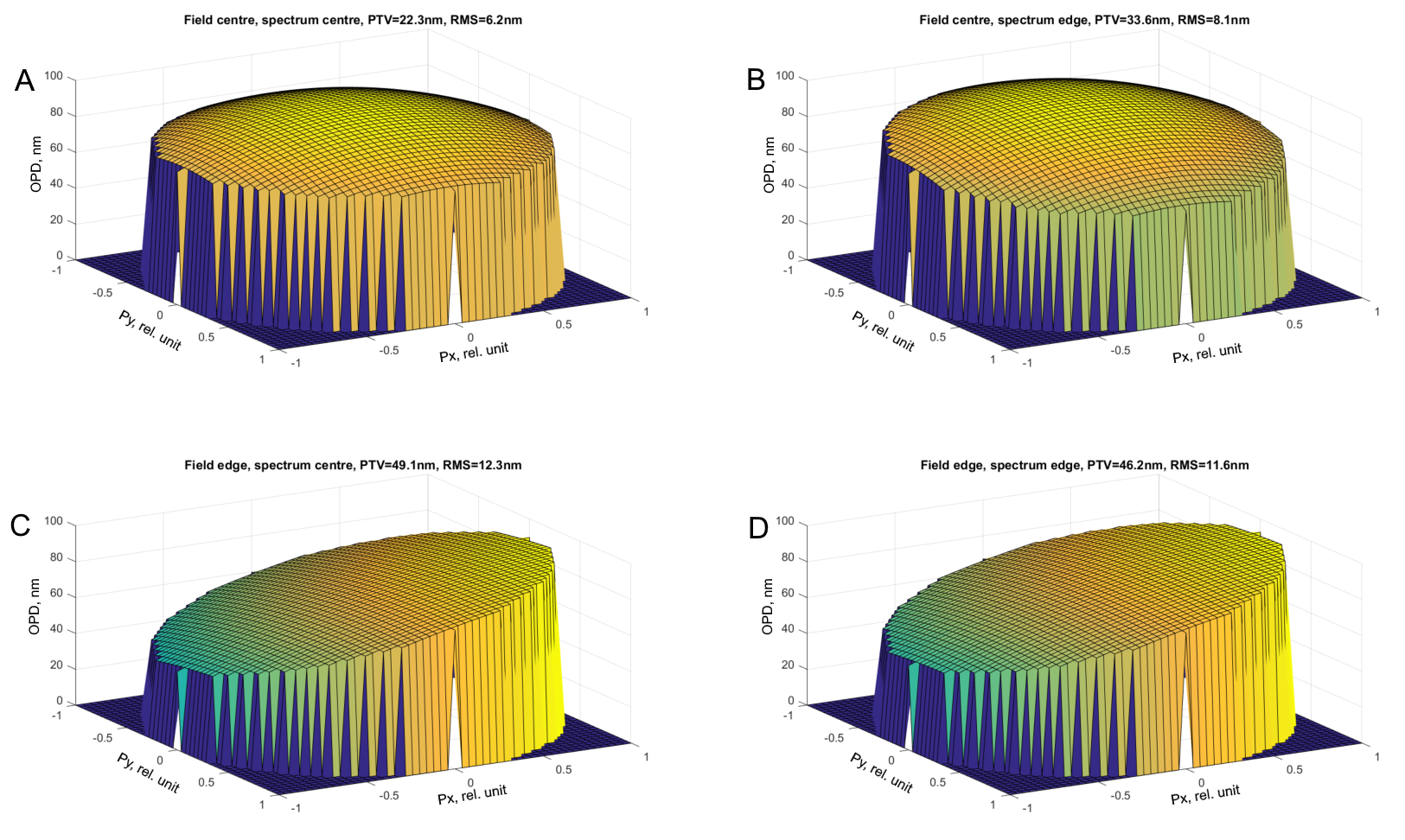}
   \end{tabular}
   \end{center}
   \caption[birefrin] 
   { \label{fig:birefrin_L2} 
Birefringence-induced WFE for ICAM L2 with LR1 grating: A -- 0 mm field, 1090 nm; B -- 0 mm field, 811 nm; C -- 270.8 mm field, 1090 nm; D -- 270.8 mm field, 811 nm. }
   \end{figure}

Stress birefringence in  optical materials may become notable for the parts of this size, but it is also difficult to account for when performing raytracing in a loop. So, we presume that the birefringence-induced WFE is just proportional to the propagation length inside the material and assign the coefficient of 10 nm/cm, which is typical for a high quality material. Then we explore the optical path distributions over the pupil, field and spectrum to estimate the effect on the overall WFE. The worst case of  ICAM L2 is given in Fig.~\ref{fig:birefrin_L2}. This analysis presumes that the total stress birefringence caused by the blank production, polishing, adhesive bonding, mounting and thermal shrinkage will not exceed the specified limit. Also, in this case it is applied as a uniform relative optical path difference across the aperture. In the reality it will be rather a random value with varying sign. Converting the results into RMS WFE and applying RSS rule for summing up individual contribution should moderate the effect of these simplifications.

   In a similar way, the refraction index inhomogeneity is difficult to introduce explicitly in the Monte-Carlo loop. It could be represented as a gradient index material, but this profile should be generated separately for each lens and plate at every wavelength. We expect that the inhomogeneity will have a rotationally symmetric distribution \cite{Jedamzik19} for lenses as  

   \begin{equation}
\label{eq:inho}
\Delta n = n_0+a_2 \rho^2+a_4 \rho^4+a_2 \rho^4.
\end{equation}

\begin{figure} [ht]
   \begin{center}
   \begin{tabular}{c} 
   \includegraphics[width=0.86\textwidth]{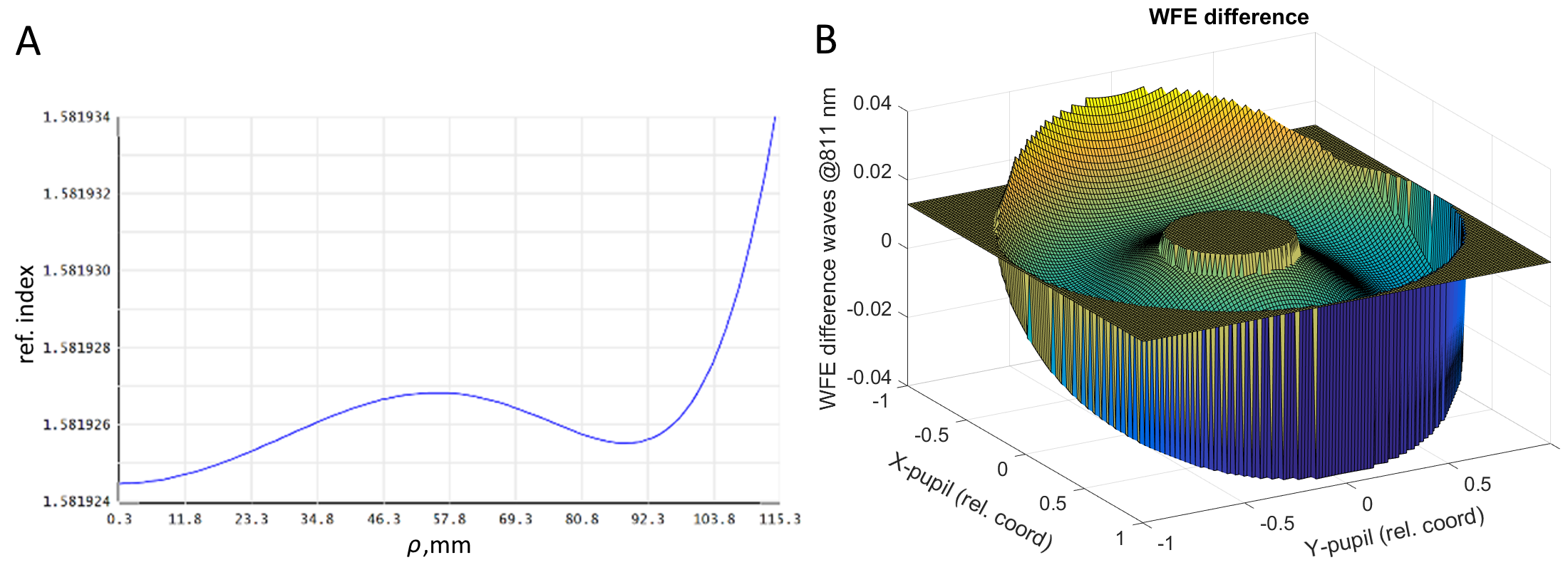}
   \end{tabular}
   \end{center}
   \caption[inhomog] 
   { \label{fig:inhomog} 
Refraction index inhomogeneity: A -- sample index distribution for ICAM L1 at 811 nm, B -- introduced WFE. }
   \end{figure}

Fig.~\ref{fig:inhomog} demonstrate the effect on example of ICAM L1 at the shortest wavelength. The inhomogeneity in this case is 10 ppm, which is a moderate value for this size and S-FTM16 glass. The corresponding additional WFE is 12.5 nm RMS. 

Another effect, which is difficult to introduce to the Monte-Carlo analysis, is the optical surface deformations under the own weight. The positioning errors under the gravitational load is a subject of purely mechanical design and analysis. The thermal deformations are compensated with low CTE materials and flextures, as it was shown by analysis, so they can be excluded from consideration in the WFE budget.

 \begin{figure} [ht]
   \begin{center}
   \begin{tabular}{c} 
   \includegraphics[width=0.8\textwidth]{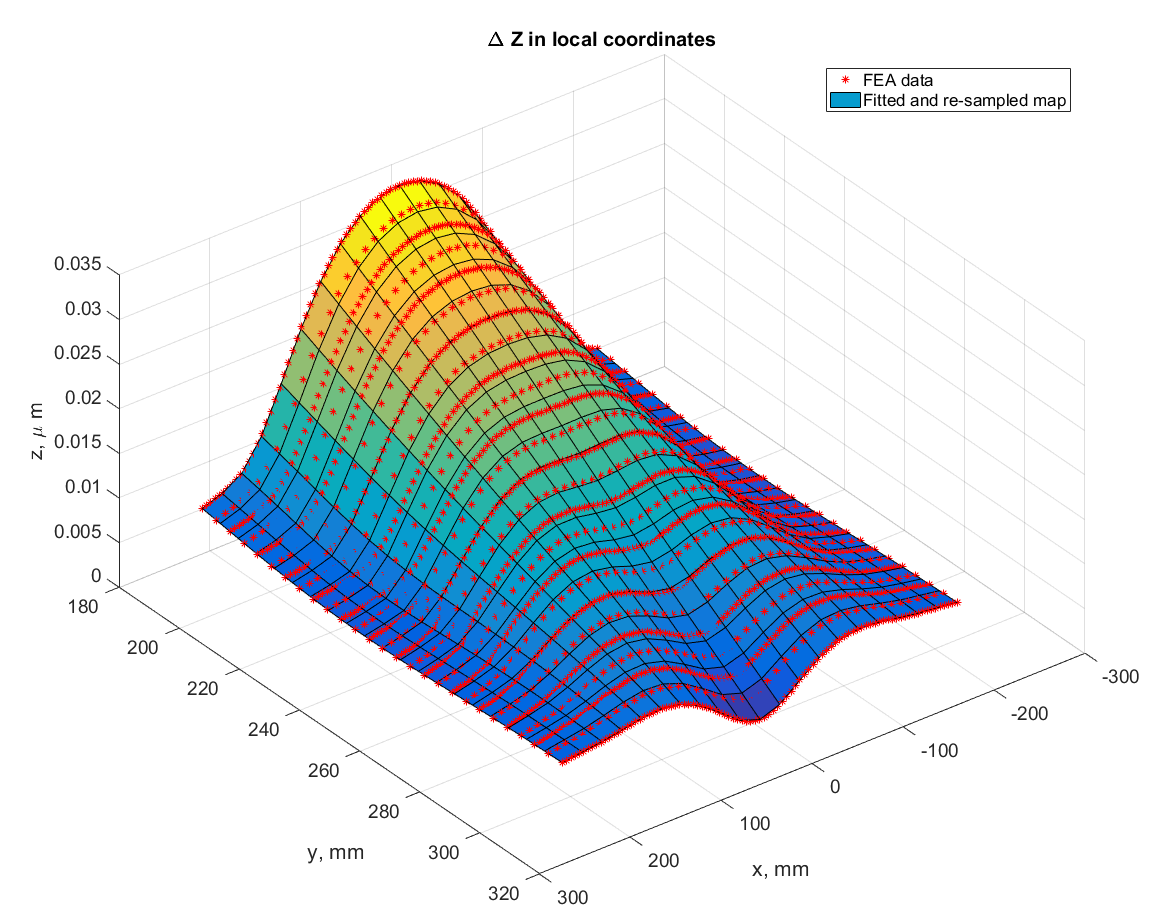}
   \end{tabular}
   \end{center}
   \caption[fea] 
   { \label{fig:fea} 
Collimator M2 optical surface deformation under own weight as an example of finite element analysis results used for the WFE calculations.}
   \end{figure} 

   However, the gravitational surface deformations are unavoidable and may be significant for the large parts. 
Therefore, they have been modelled for each surface in every position with the finite element analysis (FEA) software,  re-sampled across a regular rectangular grid and introduced to the raytracing model wit the Grid Sag tool. An example of \textit{M2} mirror deformation map is given in Fig.~\ref{fig:fea}. This is a deterministic effect, so there is no need to run a statistical analysis for it. On another hand, the corresponding error adds linearly to the nominal WFE.
 
 Error of CTE measurement may affect the aspheres shapes difference between the warm and cold conditions and should be taken into account via RSS. For the collimator mirrors this effect must be negligible due to the extremely low nominal CTE of Zerodur. However for the lenses the CTE at cryogenic temperatures are measured with uncertainty of $10^{-6} 1/K$ or $\approx12.5 \%$ \cite{Brown2004}. Being applied to the ICAM L1  asphere shape, this error corresponds to an additional WFE of 2 nm excluding defocus. 
 
\begin{figure} [ht]
   \begin{center}
   \begin{tabular}{c} 
   \includegraphics[width=0.95\textwidth]{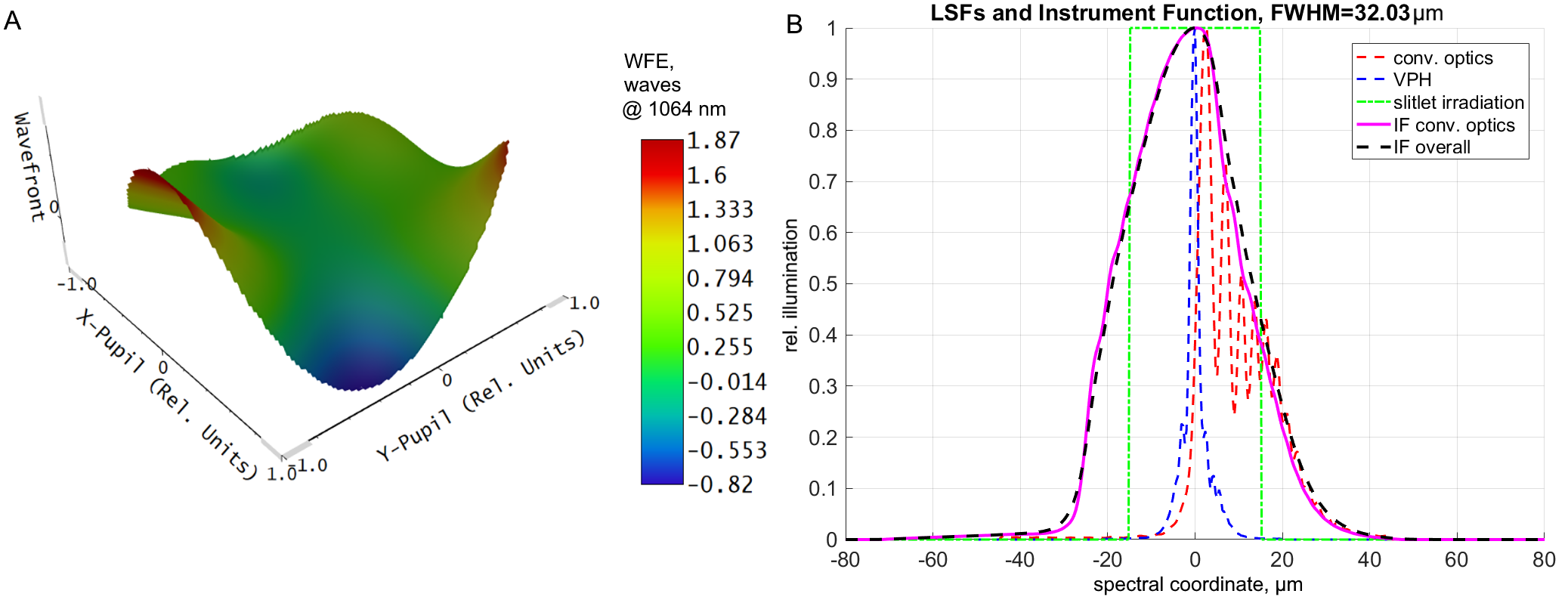}
   \end{tabular}
   \end{center}
   \caption[lsf] 
   { \label{fig:lsf} 
Instrument function computations: A -- WFE map of the VPH grating for the median case,  B -- LSF and IF plots for different contributors. }
   \end{figure}

Finally, the  VPH grating introduces a notable WFE, which occurs because of a combination of mechanisms. This includes the substrates shapes errors, dichromated gelatin (DCG) index inhomogeneity and thickness variation, those of the adhesive, local variation of the grooves pattern occuring because of imperfections of the recording wavefronts and the holographic setup optics and mechanics. We presume that the WFE introduced by the gratings remains relatively stable across the working waveband and is driven by low order Zernike modes. So, we generate a median case, using the distribution across the first 25 modes, observed in experiment (see paper by E. Meyer et al. at this conference). This WFE map is shown in Fig.~\ref{fig:lsf}A. Then we generate 100 instances for the grating only, presuming that the WFE is normally distributed around the median with a $99.7\%$  probability ($\pm 3 \sigma $ equivalent) to be below 949 nm RMS -- or $1.5 \lambda$ at control wavelength of 633 nm. We scale it to meet the $95.4\%$ probability criterion used for other statistical esimates and obtain 727 nm WFE. These values are defined for the full clear aperture. For other spaxel scales they are scaled by the clear aperture areas ratio.

So, all of the effects discussed above are summarized in Table~\ref{tab:wfeSum}. The WFE errors are categorized by the physical mechanism and contributing element or module, where possible.
  
   \begin{table}[!ht]
\caption{As-built WFE summary in nanometers} 
\label{tab:wfeSum}
\begin{center}       
\begin{tabular}{|p{2cm}|p{2cm}|c|c|c|c|} 
\hline
Mechanism &	Element/ module	& 60x30 mas	& 20x20 mas &	10x10 mas & 	4x4 mas \\
\hline
Monte-Carlo &	Convent. optics &	320.18 & 212.40 &	73.66	& 33.07\\
\hline
\thead{Gravitation\\ deformation} &	Collim	& 11.23 &	4.37&	1.12&	0.52\\
\hline
                &	FMM&	26.1&	13.3&	5&	2.3\\
\hline
                &	VPH&	0.1&	0.1&	0&	0\\
\hline
            &ICAM&	0.5&	0.3&	0&	0.1\\
\hline
            &  Overall &	37.93&	18.07&	6.12&	2.92\\
\hline
\thead{Mid-freq. \\surface errors} &	M1&	82.7&	22.99&	6.86	&2.37\\
\hline
                            &	M2&	72.4&	20.13&	6.01&	2.39\\
\hline
                &	M3&	87.1&	24.21&	7.23&	2.87\\
\hline
                &	FM1&	14.7&	4.09&	1.22&	0.49\\
\hline	
            &FM2	&9.1	&2.53&	0.76&	0.30\\
 \hline     
                &	ICAM&	15.0&	4.17&	1.25&	0.50\\
\hline
Birefring.&	L1&	5.9&	1.64&	0.49&	0.19\\
	\hline
 &L2	&12.3&	3.42&	1.02&	0.41\\
\hline
&L3	&11.7&	3.25&	0.97&	0.39\\
\hline
&L4	&0.9&	0.25&	0.07&	0.03\\
\hline
&L5	&0.8&	0.22&	0.07&	0.03\\
\hline
Inhomog.	&ICAM&	27.73 &	7.71&	2.30&	0.91\\
\hline
CTE error &	ICAM&	3.0	&0.83	&0.25	&0.10\\
\hline
Hologram &VPH	&727.00	&202.11&	60.34&	23.99\\
\hline
Total without grating& &	386.70& 234.01&	80.69&	36.31\\
\hline
Total with grating& &823.45	&309.21&	100.76&	43.52\\
\hline

\end{tabular}
\end{center}
\end{table} 

With the current approach and the WFE of individual elements and modules provided by the potential manufacturers we see a major impact of the VPH grating WFE on the overall wavefront error budget. This may become the main driver for the sub-system manufacturing. One of possible options to reduce this effect is using Magnetorheological Finishing (MRF) on the grating substrate to introduce a correcting wavefront deformation. This solution is currently in active development (see paper by E. Pearson et al. at the same conference). 

Also, as it was stated in the beginning, the WFE criterion serves as an intermediate metric, which is easier to use for breaking down the image quality budget and controlling in practice. However, FWHM and spectral resolving power criteria represent better the main function of the spectrograph. To analyze its' statistical behaviour we performed a parallel Monte-Carlo simulation for a system, including a perfect collimator and camera and a VPH grating with WFE represented by the last surface error controlled by the low order Zernike modes $Z_{4-25}$. Then for each of the instances in conventional optics and VPH models we found the line spread function and convolved it with the rectangular function, representing the slitlet luminosity:

 \begin{equation}
\label{eq:conv}
E_{ISP}=(RECT\circledast LSF_{conv.opt.})\circledast LSF_{VPH},
\end{equation}

then the spectral resolving power is computed taking into account the measured FWHM of this instrument function and the actual reciprocal linear dispersion

\begin{equation}
\label{eq:R}
R=\frac{\lambda}{\frac{d \lambda}{ d l} FWHM}.
\end{equation}

\begin{figure} [ht]
   \begin{center}
   \begin{tabular}{c} 
   \includegraphics[width=0.95\textwidth]{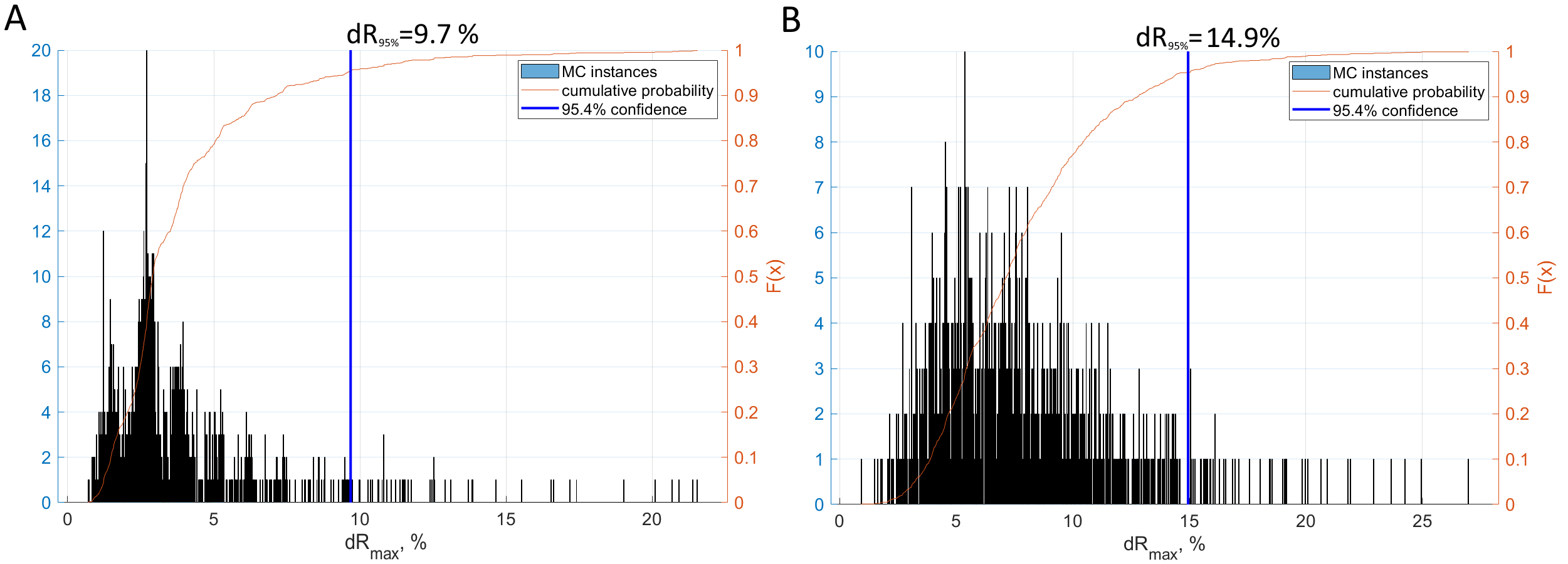}
   \end{tabular}
   \end{center}
   \caption[res] 
   { \label{fig:res} 
Monte-Carlo statistics for the spectral resolving power degradation from the nominal value over all the working configurations: A -- conventional optics only, B -- including VPH.}
   \end{figure}

The results of spectral resolving power statistical analysis are shown in Fig.~\ref{fig:res} as a degradation from the nominal value in percents. As it was shown above with the WFE metric, meeting the requirement, which corresponds to $\leq10\%$ in this case, strongly depends on the VPH wavefront error. However, there is no direct correlation between the two metrics. Another features to note in these plots are their high asymmetry and presence of local peaks associated with individual gratings. Although, the latter ones are significantly blurred by the VPH WFE factor.

\section{Conclusion and future work}
\label{sec:conc}  

The present analysis demonstrates that the current HARMONI spectrograph sub-system design can reach the required performance in terms of the image quality with all of the manufacturing and alignment errors acting simultaneously. Individual tolerances may be challenging, but all of them are reachable with the current technological level as confirmed by the suppliers. With a $95.4\%$ confidence we expect the 95-percentile WFE to be below 823.45, 309.21, 100.76 and 43.52 nm for the 4 spaxels scales. In a case of applying MRF correction for the grating wavefront we may hope to reduce these values to the levels of $\approx$ 393, 235,	81	and 36 nm, respectively. The expected spectral resolution degradation is $14.9\%$.
The outcomes of this analysis will be used in the final specifications of optical parts and modules for HARMONI ISP.

We present an analysis sequence, which combines finding the numerical derivatives and applying RSS rule for definition of the initial tolerances, their correction to a practically reachable level and statistical analysis of as-built system performance relying on Monte-Carlo simulations in a few parallel models. We hope that this approach may be useful for other engineers and researchers, who face a problem of tolerance analysis in optical system with multiple levels of functionality break-down, complex requirements and uncertainty introduced by the technological restrictions.

\section*{ACKNOWLEDGMENTS}       
This work is supported by UKRI-STFC grants $\# ST/X002322/1$ and $ST/S001409/1$.  Authors also acknowledge support from a philanthropic donation from Philip and Roswitha Wetton towards the LR2 development.

\bibliography{main}
\bibliographystyle{ieeetr}

\end{document}